\documentclass[runningheads]{llncs}
\usepackage[T1]{fontenc}
\usepackage{hyperref}
\usepackage{graphicx}
\begin{document}
\title{Experiences Applying Lean R\&D in Industry-Academia Collaboration Projects}
\titlerunning{Experiences Applying Lean R\&D}
%
\author{Marcos Kalinowski \orcidID{0000-0003-1445-3425}
\\
Lucas Romao\orcidID{0009-0001-2153-1498}
\\
Ariane Rodrigues\orcidID{0000-0002-1614-918X}
\\
Clarissa Barbosa\orcidID{0009-0006-9105-7062}
\\
Hugo Villamizar\orcidID{0000-0003-4142-6967}
\\
Simone~D.~J.~Barbosa\orcidID{0000-0002-0044-503X}
\\
Hélio~Lopes\orcidID{0000-0003-4584-1455}}

\authorrunning{Kalinowski et al.}
%
\institute{
Pontifical Catholic University of Rio de Janeiro (PUC-Rio), Rio de Janeiro RJ 22451-900, Brazil \\
\vspace{3mm}
\email{\{kalinowski, lromao, arodrigues, cbarbosa, hvillamizar, simone, lopes\}@inf.puc-rio.br} \\
\vspace{3mm}
}
\maketitle             
\begin{abstract}
Lean R\&D has been used at PUC-Rio to foster industry-academia collaboration in innovation projects across multiple sectors. This industrial experience paper describes recent experiences and evaluation results from applying Lean R\&D in partnership with Petrobras in the oil and gas sector and Americanas in retail. The findings highlight Lean R\&D’s effectiveness in transforming ideas into meaningful business outcomes. Based on responses from 57 participants—including team members, managers, and sponsors—the assessment indicates that stakeholders find the structured phases of Lean R\&D well-suited to innovation projects and endorse the approach. Although acknowledging that successful collaboration relies on various factors, this industrial experience positions Lean R\&D as a promising framework for industry-academia projects focused on achieving rapid, impactful results for industry partners.

\end{abstract}
\section{Introduction}

Industry-Academia Collaboration (IAC) projects have been a main focus at PUC-Rio, Latin America's leading university in industry income according to the Times Higher Education ranking \cite{latinAmericaRankings2023}. Several of these IAC projects are carried out within the oil and gas industry, mainly motivated by the needs of Petrobras, a state-owned Brazilian multinational corporation headquartered in Rio de Janeiro, Brazil.

Research, Development, and Innovation (RDI) initiatives in the oil and gas industry commonly involve cooperation terms with research institutes and universities. These terms were usually designed in a plan-driven manner, with deliveries that, given research uncertainties, typically spanned several years. In September 2019, representatives of Petrobras came up with a challenge: to structure an RDI collaboration that would enable agile deliveries of Minimal Viable Products (MVPs) to test business hypotheses, allowing to take ideas from their conception to impact.

To address this challenge, we designed an RDI approach to allow fast-paced MVP deliveries. The resulting approach, inspired by continuous software engineering principles \cite{fitzgerald2017continuous} and concepts from Lean \cite{monden2011toyota} and Lean Startup \cite{ries2011lean}, was called Lean R\&D \cite{kalinowski2020towards,kalinowski2020lean}. In January 2020, Petrobras started four IAC projects with PUC-Rio applying Lean R\&D. The ExACTa (\href{http://exacta.inf.puc-rio.br}{exacta.inf.puc-rio.br}) laboratory was created at PUC-Rio to handle such agile RDI projects, aiming at fast-paced deliveries of R\&D projects, aligning research on artificial intelligence, human-computer interaction, and software engineering with the digital transformation needs of industrial partners.

The first case studies on using Lean R\&D provided initial indications into the feasibility of using the approach within RDI projects \cite{kalinowski2020lean,teixeira2021lessons}. Since then, ExACTa has been using Lean R\&D to deliver innovative solutions to several industry partners. This industrial experience paper provides new results assessing the use of Lean R\&D in more recent RDI IAC projects with Petrobras and Americanas, a player active in the retail business in Brazil, which owns more than 1,000 physical stores spread throughout the country. We report on the experiences and their results, assess the perception of the suitability of Lean R\&D, and discuss the overall acceptance from the point of view of members of the R\&D teams, business and technical managers, and sponsors. 

With Petrobras, we focus on the use of Lean R\&D in two Artificial Intelligence (AI) projects. Both were subject to patent applications \cite{kuramoto2023methodSmartTocha,kuramoto2023methodDigitalInspector} and received a Petrobras inventor award at the end of 2022. One of these projects, called Smart Tocha (\textit{Smart Torch}), comprises an AI solution that continuously processes oil refinery torch burn images to automatically act over refinery controls to increase the energy efficiency of the gas burning. The other one, called Inspetor Digital (\textit{Digital Inspector}), involves using AI to help refinery equipment inspectors complete reports more quickly and efficiently. This paper focuses on the process used at ExACTa, not on the particular projects; more details on the projects can be found within the patent applications.

Americanas started their IAC with ExACTa in July 2022 and set up four teams to simultaneously handle RDI projects using Lean R\&D. These projects concerned innovating the management of physical stores, improving logistic operations, and applying machine learning for product classification and e-commerce recommendations. While ExACTa R\&D employees working on the projects for Petrobras were experienced professionals, Americanas wanted to involve undergrad, master, and Ph.D. students in the teams. The contract involved applying two complete Lean R\&D cycles throughout one year (six months for each MVP delivery per team).

Each team had ten members: eight undergraduate students, one master's student, and one PhD student. The team was supported by a general coordination involving representatives of ExACTa and Americanas and a dedicated mentor with a Ph.D. in computer science. The students also received training and mentorship from researchers of PUC-Rio. We called these teams ExACTa FIT, where FIT is a Portuguese acronym standing for education in technological innovation. The ExACTa FIT teams conducted all Lean R\&D phases, including ideating innovative solutions.

We report on the outcomes of these projects and perceptions on Lean R\&D gathered through questionnaires applied to the different stakeholders at Petrobras and Americanas. We received 57 responses, mainly indicating the suitability of the Lean R\&D's phases and its overall acceptance, but also providing attention points related to the ease of learning and applying the approach and improvement opportunities.

\section{Overview of Lean R\&D}

When designing the Lean R\&D approach, the main requirements were to allow maximizing business value while minimizing waste, allowing to fail fast to enable handling opportunities, and enabling to address complex RDI problems with fast-paced deliveries. 

\begin{figure}[h!]
\label{leanRD}
\centerline{\includegraphics[width=\linewidth]{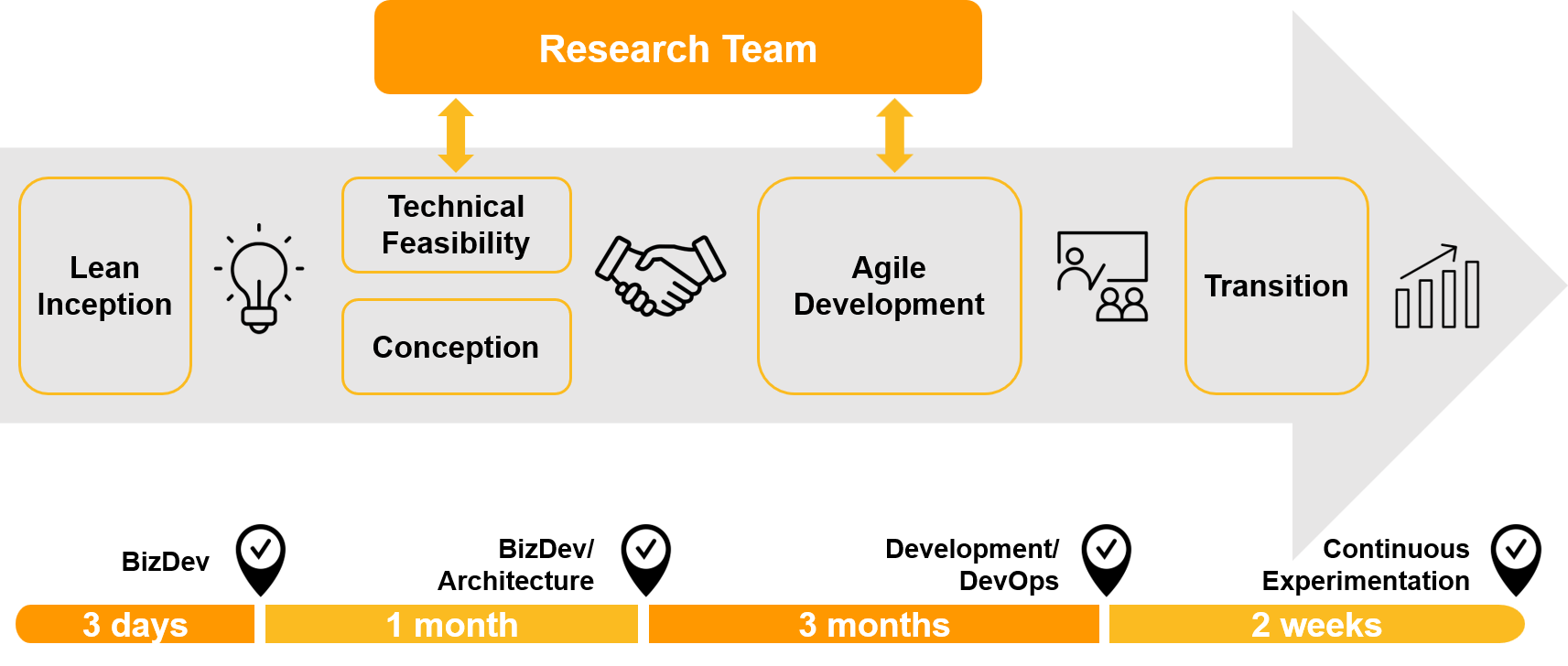}}
\caption{Overview of the Lean R\&D approach. The timeline is illustrative and reflects typical timeframes used within the ExACTa PUC-Rio laboratory.}
\label{leanrd}
\end{figure}

Based on these requirements, the approach was designed starting with a Lean Inception \cite{caroli2018lean} ideation workshop, parallel (early) technical feasibility assessment and conception phases, scrum-based development, and continuous experimentation. An overview of Lean R\&D is shown in Figure \ref{leanrd}.

It is possible to observe four checkpoints, a set of activities, and the support of a dedicated research team for technical solution-related activities. More details on these elements and the involved roles can be found in our previous paper describing Lean R\&D \cite{kalinowski2020lean}. Hereafter, we summarise the activities to provide an overview understanding.

The approach starts with a \textit{Lean Inception} to allow stakeholders to jointly outline the vision of an MVP that can be used to test business hypotheses associated with industrial impact. This phase emphasizes co-creation between academic and industry teams, ensuring alignment on the MVP's objectives and expected outcomes.

At the first checkpoint, a steering committee (including the sponsor customer) must approve the MVP outline and its intended business value. Otherwise, if the idea of the MVP or its expected results are not promising enough, it fails fast, and a new Lean Inception is conducted, potentially focusing on a different business problem. This checkpoint thus ensures that only high-potential projects move forward, optimizing resource allocation and focus.

In the \textit{Technical Feasibility} phase, the development team, assisted by the research team, starts investigating the technical feasibility of implementing the features identified during the Lean Inception. In parallel, the \textit{Conception} involves a Product Owner (PO) detailing the MVP features identified during the Lean Inception by applying product backlog-building co-creation dynamics and other typical agile requirements elicitation and specification techniques (e.g., user stories with acceptance criteria). A UX/UI designer also details the MVP’s user experience and user interface in this phase, ensuring a clear design direction.

The second checkpoint involves the steering committee analyzing the requirements specification, mainly based on the UI prototype and the results of the technical feasibility assessment, to decide whether the MVP should proceed to development. This checkpoint helps to prevent investments in technically challenging solutions that may not yield the desired business outcomes.

Thereafter, the \textit{Agile Development} phase involves the development team, with the support of the research team, implementing the MVP following Scrum ceremonies. Customer representatives are expected to actively participate in the sprint planning and reviews, providing continuous feedback and helping the team to align closely with the end user's needs and expectations. 

Following development, the third checkpoint requires the PO to present the MVP to the steering committee, where they decide on its readiness for production transition. This ensures that the MVP meets the defined criteria for value and functionality before entering the market.

Finally, the \textit{Transition} phase involves the development and infrastructure team preparing the MVP for beta testing in its final environment, where continuous experimentation approaches are applied to assess the MVP’s effectiveness in confirming the business hypotheses. 

The last checkpoint focuses on evaluating continuous experimentation results to determine whether the business hypotheses were validated and if future investment in additional MVP increments is justified. This data-driven approach helps in making informed decisions for scaling or adapting the MVP based on real-world outcomes.

\section{Applying Lean R\&D}

We analyze the application of Lean R\&D to characterize the approach with respect to its suitability for RDI IAC projects and its general acceptance from the point of view of the teams, the business and technical managers from the customer side, and customer sponsors. Beyond describing the application outcomes, we applied a questionnaire to the different stakeholders to support our analyses, with questions focusing on the suitability of the Lean R\&D phases and the overall approach acceptance. The questionnaires and anonymized raw data can be found at \url{https://doi.org/10.5281/zenodo.8317454}.

From the two projects with Petrobras, we received a total of 15 responses (out of 19 invitations): twelve from the ExACTa R\&D team members (ten from ExACta and two from Petrobras working as POs), two from technical managers of Petrobras, and one from the main sponsor of the artificial intelligence in refinery operation projects. They were highly experienced, with a minimum experience of four years and a median of eight years, and covered all the roles involved in the Lean R\&D approach in both projects. The technical managers and sponsor answered separate questionnaires more objectively focused on the method's acceptance.

From the projects with Americanas, we received a total of 42 responses (out of 48 invitations): 38 from the ExACTa FIT team members, one from a technical manager of Americanas, one from a business manager of Americanas, and two from the sponsors responsible for Americanas' digital innovation projects. They were much less experienced (students, including MS and Ph.D. students), with a minimum and median experience of one year. They also cover the roles involved in the Lean R\&D approach. As done for Petrobras, again the technical managers and sponsors answered a separate questionnaire focused on the method's acceptance.

\section{Lessons Learned}

\subsection{Lean R\&D Application Outcomes}
In both experiences, Lean R\&D enabled defining joined MVP visions, addressing research-related uncertainties early (in the technical feasibility phase), and efficiently delivering valuable MVPs. The two projects delivered to Petrobras resulted in international patent applications \cite{kuramoto2023methodSmartTocha,kuramoto2023methodDigitalInspector} and received a Petrobras inventor award at the end of 2022. 

The Smart Tocha project, according to continuous experimentation measurements, results in energy savings comparable to the energy consumption of a city with 20,000 inhabitants per refinery. Petrobras mainly uses the SAFe \cite{scaledAgileFramework} scaled agile framework for software development projects, and the Smart Tocha MVP was loaded into a SAFe agile release train with the mission of rolling it out to several refineries. It is noteworthy that Lean R\&D aims to deliver innovative MVPs, not to sustain or evolve them, for which other approaches (such as SAFe) may be employed. 

The Digital Inspector MVP was also successfully delivered. It was integrated into Petrobras' equipment inspection system and, based on continuous experimentation measurements, is helping refinery equipment inspectors to complete reports more quickly and efficiently.

On the overall suitability of Lean R\&D for innovation projects, the sponsor at Petrobras mentioned his satisfaction with the fast-paced deliveries and business results in the questionnaire: ``Lean R\&D is appropriate for research and development innovation projects; it sped up the deliveries and business results. Furthermore, it avoids wasting effort on projects that are not promising."

With respect to Americanas, the MVPs were delivered as scheduled in January and July 2023 and are also positively impacting the business. Unfortunately, we have no continuous experimentation results at hand yet. One of the sponsors answered that ``Lean R\&D was very appropriate for the projects, not just because it made them agile, but mainly because it placed all participants as protagonists in understanding the problem, ideating the solution and building the MVP, fundamental steps to favor innovative solutions." The other sponsor answered ``The Lean R\&D approach helped to structure the innovation projects, from conception to development, providing a methodology for integrating the business areas and the research and development teams." 

The satisfaction with the project can also be perceived in a \href{https://www.linkedin.com/posts/americanas-sa_americanas-futuro-lab-puc-rio-activity-7100866017156841472-Nvmo}{public communication} made by the company in its LinkedIn profile (with more than one million followers): ``The project aimed to allow students to find solutions to real-world problems and enrich their training, combining it with graduate-level research and direct exposure to the company's specific challenges. The projects were carried out with great efficiency and dedication by the students and the professors from PUC-Rio in partnership with the business areas and the digital innovation area of Americanas."

\subsection{Suitability of the Lean R\&D Phases}
\begin{figure}[h!]
\centerline{\includegraphics[width=\linewidth]{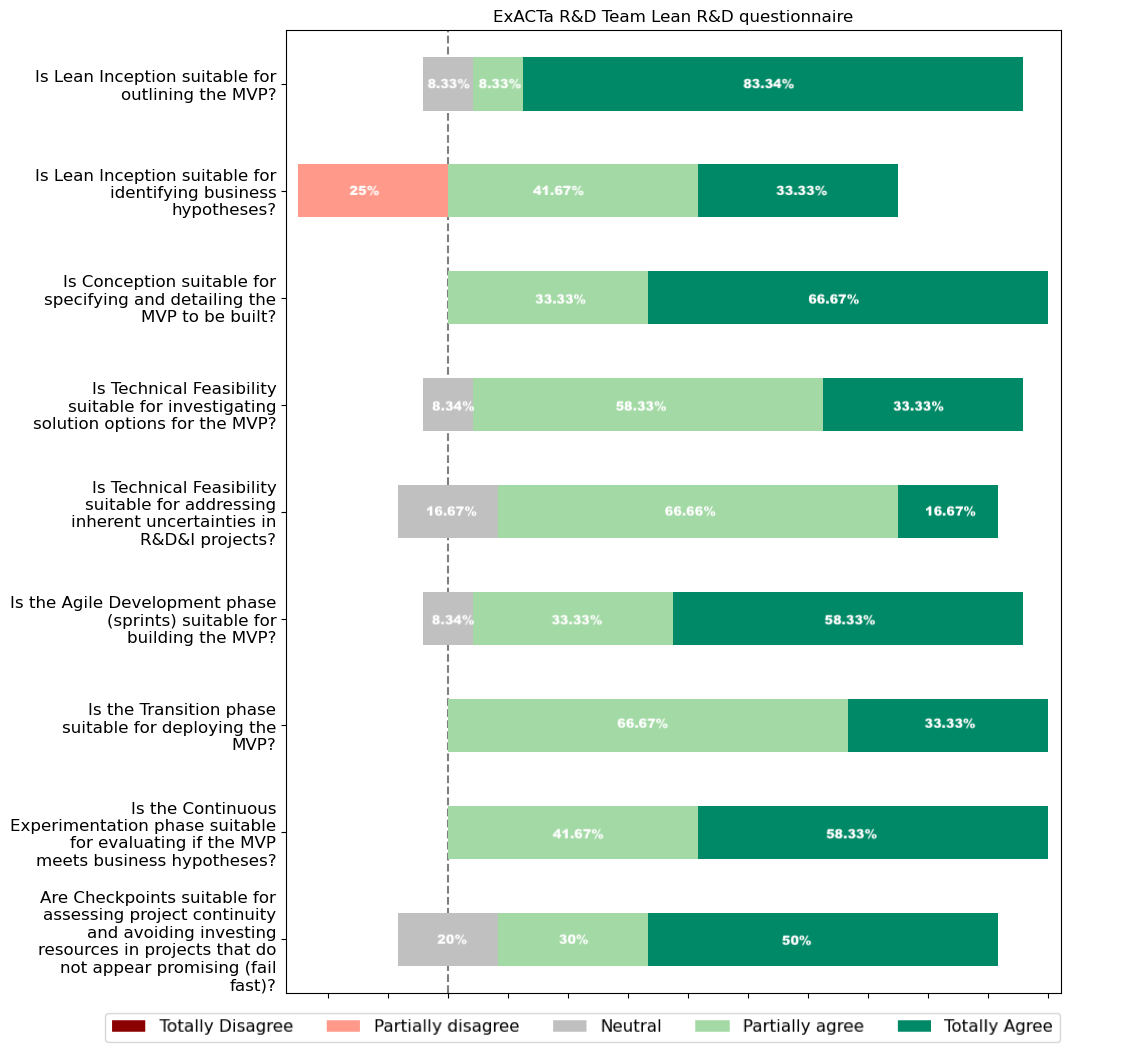}}
\caption{Suitability of Lean R\&D phases according to the ExACTa R\&D team working with Petrobras.}
\label{phaserd}
\end{figure}

\begin{figure}[h!]
\centerline{\includegraphics[width=\linewidth]{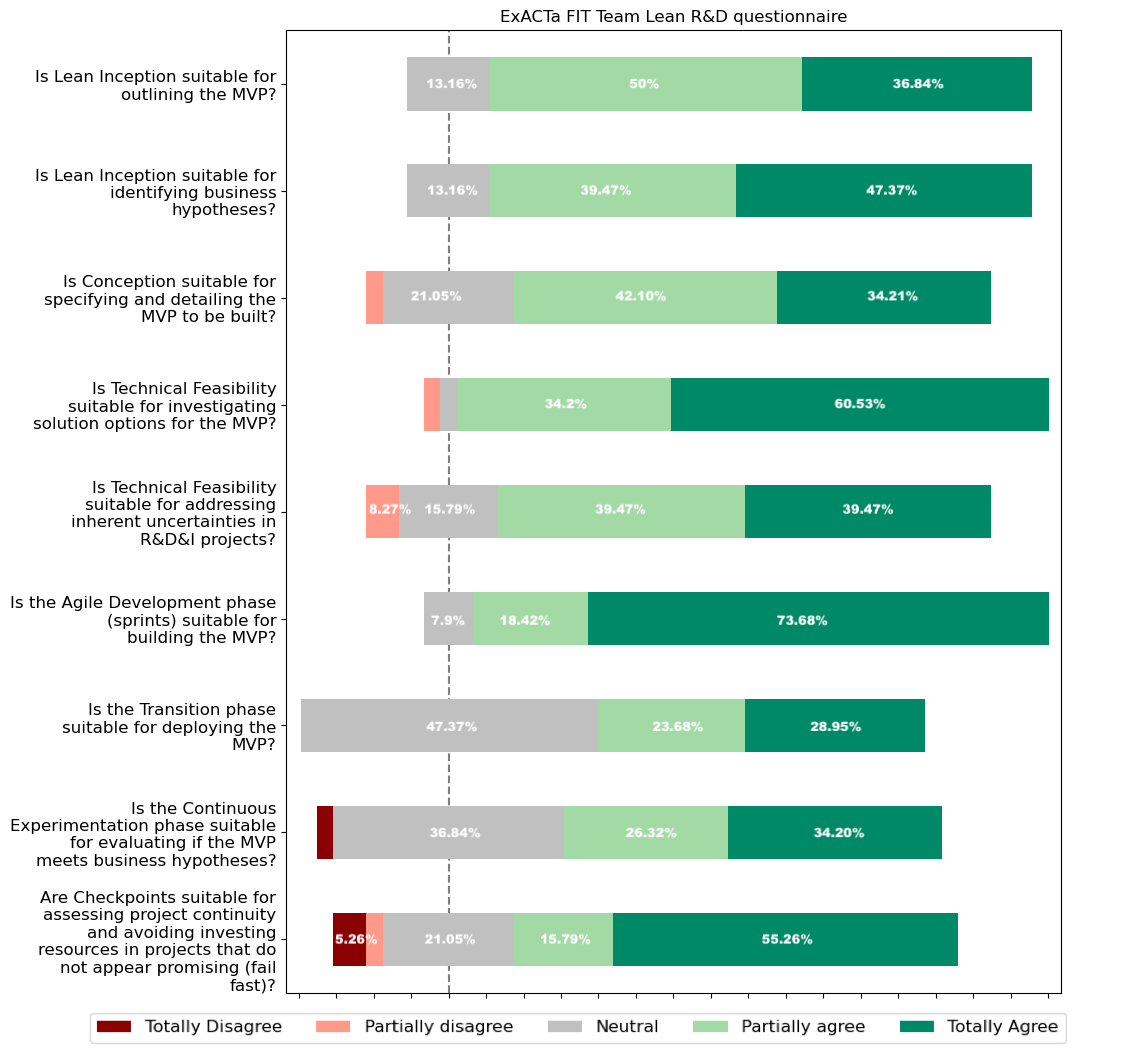}}
\caption{Suitability of Lean R\&D phases according to the ExACTa FIT team working with Americanas.}
\label{phasefit}
\end{figure}

We analyzed the answers of the Petrobras and Americanas teams regarding their agreement on the suitability of each phase with respect to its purpose. We used a five-point Likert scale (totally disagree, partially disagree, neutral, partially agree, and totally agree). The results for the ExACTa R\&D team working with Petrobras and the ExACTa FIT team working with Americanas are shown in Figures \ref{phaserd} and \ref{phasefit}. 

As shown in Figure \ref{phaserd}, in the more experienced R\&D team working with Petrobras, three out of twelve team members partially disagreed that Lean Inceptions are suitable for identifying business hypotheses. Two of them justified their answers. They explained that Lean Inception is concerned with outlining the MVP and describing related business hypotheses but not identifying them, which might require analyzing the business. Indeed, they precisely highlighted that the main purpose of Lean Inception is focused on outlining the MVP to deliver business value.

As shown in Figure \ref{phasefit}, in the team working with Americanas, one of the 38 participants partially disagreed on the suitability of the conception phase for detailing the MVP. However, the justification only pointed to difficulties in conducting the conception phase. This same participant also partially disagreed on the technical feasibility, together with one other, who questioned the support of this phase for addressing uncertainties. Based on their comments, they were actually referring to problems within their team, which did not get sufficient stakeholder involvement. Furthermore, one participant disagreed on the suitability of the continuous experimentation phase. Mainly, this participant was worried about end-users using an MVP with an extremely limited set of features. While this might indeed be a risk, this aspect forces the focus on the main features to add value. 

Finally, two participants disagreed on the importance of the checkpoints, and one partially disagreed. However, based on their justifications, they did not understand the purpose of these checkpoints. One of them mentioned that the checkpoints are insufficient to communicate all project-related problems. The other one mentioned that changes can happen at every sprint planning. Discussing strategic innovation decisions in every sprint is typically not feasible.

Both figures mainly indicate an agreement on the suitability of the Lean R\&D phases from the point of view of the team members. Indeed, the teams could use the approach to innovate and deliver value. Nevertheless, we identified opportunities for improvement concerning better supporting describing business hypotheses and providing clearer expectations on the activities conducted within each phase.

\subsection{Lean R\&D Acceptance}
We elaborated statements on the perceived usefulness, ease of use, and behavioral intention to adopt Lean R\&D based on the Technology Acceptance Model and asked participants to answer according to their agreement. The questionnaire was applied to the team members and the business and technical managers and sponsors of Petrobras and Americanas. 

\begin{figure*}[ht]
\centerline{\includegraphics[width=\linewidth]{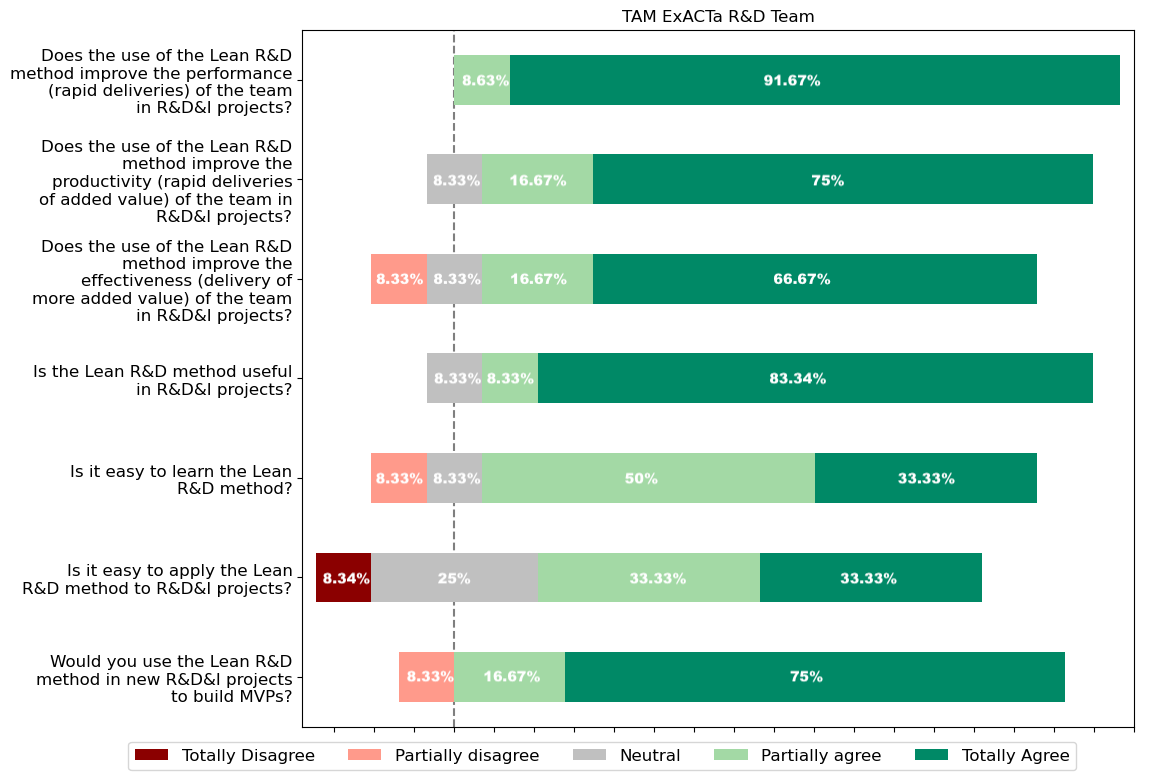}}
\caption{Technology acceptance by the ExACTa R\&D team members of the projects with Petrobras.}
\label{tamrd1}
\end{figure*}

\begin{figure*}[ht]
\centerline{\includegraphics[width=\linewidth]{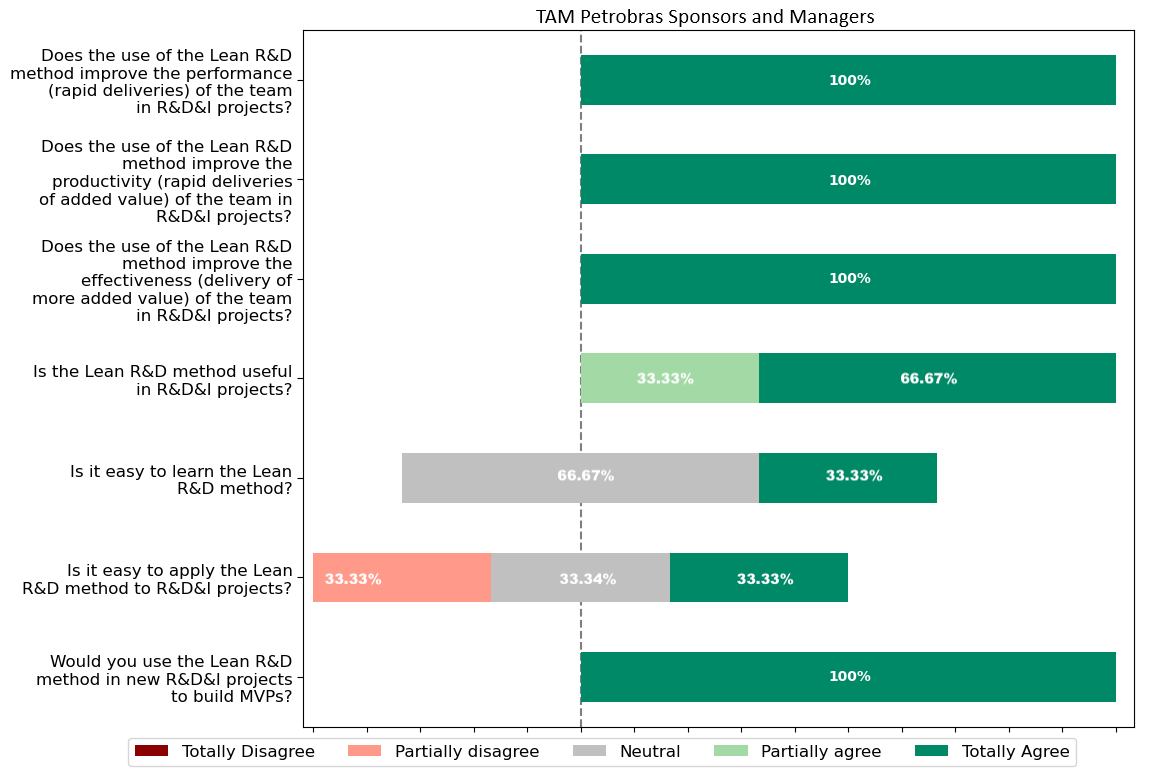}}
\caption{Technology acceptance by the business and technical managers and sponsors of Petrobras.}
\label{tamrd2}
\end{figure*}

The results for the R\&D team members can be seen in Figure \ref{tamrd1}, while the results for the managers and sponsors of Petrobras can be seen in Figure \ref{tamrd2}.

Analyzing Figure \ref{tamrd1}, it is possible to observe a clear acceptance of Lean R\&D among the R\&D team members. Out of the twelve team members working with Petrobras, one partially disagreed on the improved effectiveness (added value) and the intention to use. This participant argued that the assessment of added value should receive more attention. Nevertheless, Lean R\&D proposes continuous experimentation to measure the added value based on explicit business hypotheses. Another participant was skeptical about the ease of learning and applying Lean R\&D, but provided no additional comments. 

Based on Figure \ref{tamrd2}, it is possible to understand that the managers and sponsors at Petrobras also had a large agreement. The only partial disagreement was of the sponsor and on the ease of use. In fact, he was only involved in the checkpoints.

\begin{figure*}[ht]
\centerline{\includegraphics[width=\linewidth]{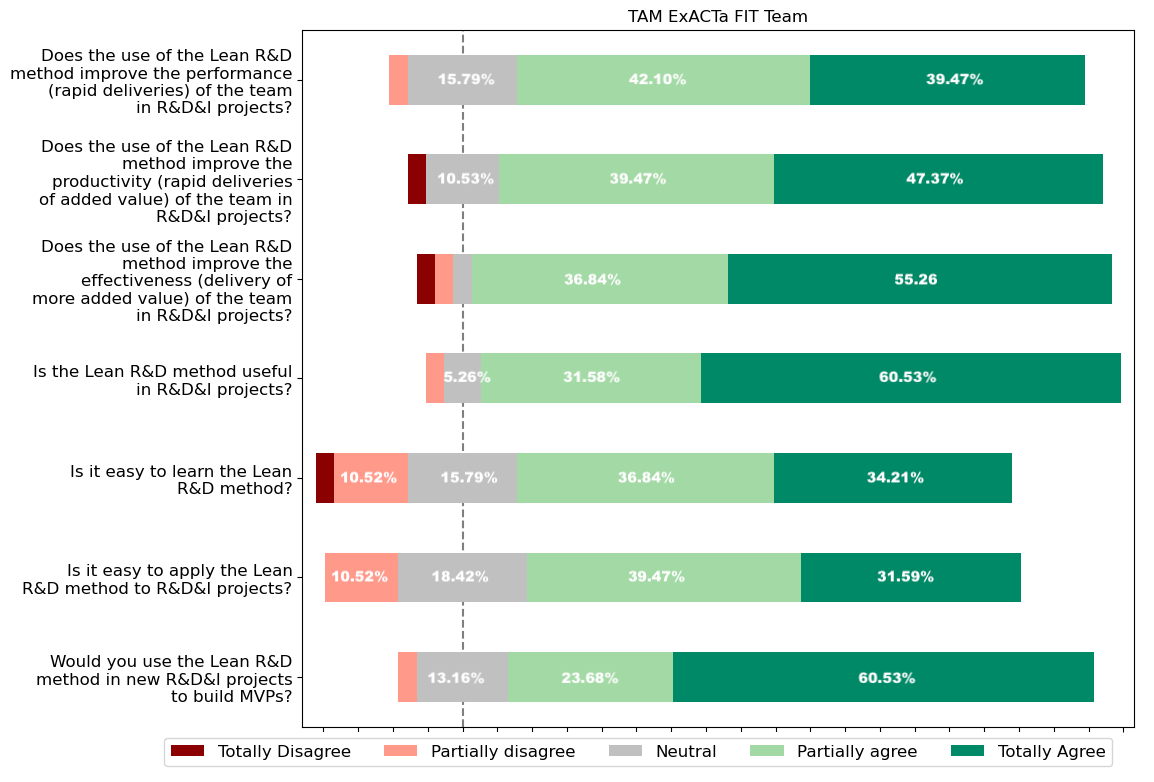}}
\caption{Technology acceptance by the ExACTa FIT team members of the projects with Americanas.}
\label{tamfit1}
\end{figure*}

\begin{figure*}[ht]
\centerline{\includegraphics[width=\linewidth]{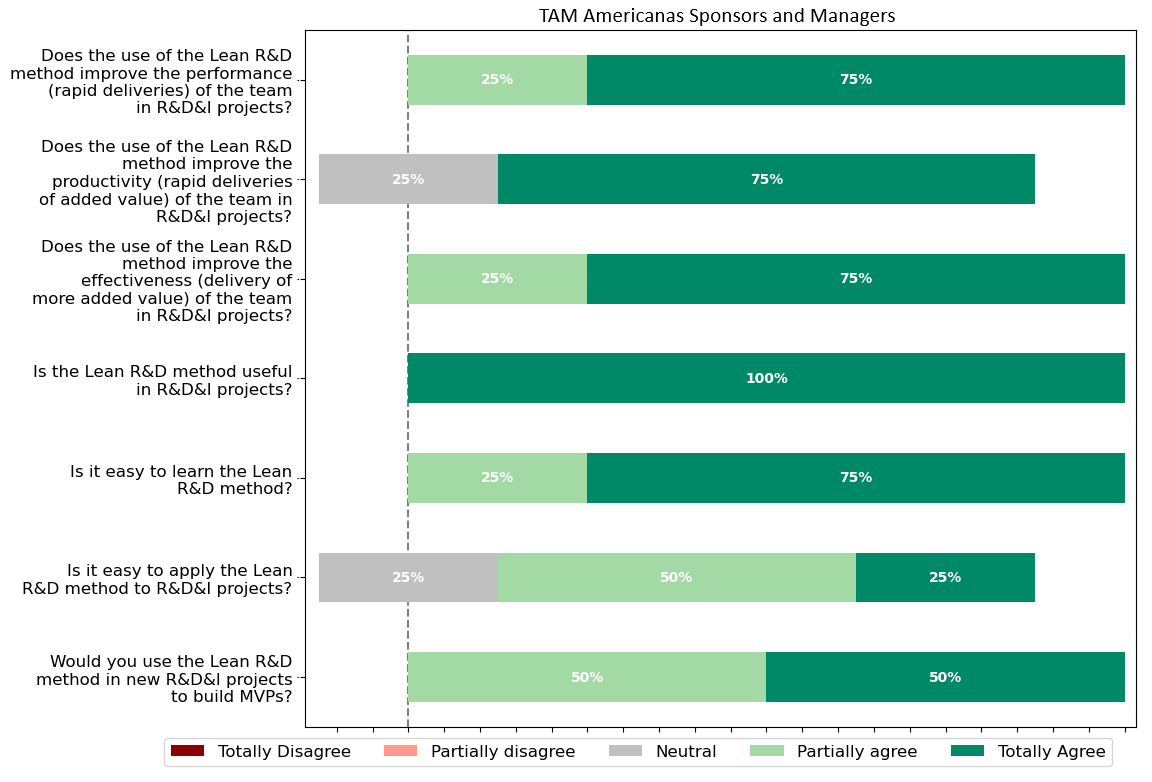}}
\caption{Technology acceptance by the business and technical managers and sponsors of Americanas.}
\label{tamfit2}
\end{figure*}

The results for the ExACTa FIT team members can be seen in Figure \ref{tamfit1}, while the results for the managers and sponsors of Americanas can be seen in Figure \ref{tamfit2}.

Figure \ref{tamfit1} shows that out of the 38 team members working with Americanas, just six partially or totally disagreed on any of the items. Only one of them commented, mentioning that applying the approach was not easy as they had limited prior experience. In fact, the most questioned statements concerned the ease of learning and applying Lean R\&D. It is noteworthy that the team members were less experienced. 

Figure \ref{tamfit2} allows to observe that the managers and sponsors at Americanas also largely agreed with the TAM statements. One of the two managers more conservatively reported being neutral on the ease of use, and one of the two sponsors reported being neutral regarding improved productivity.

\section{Discussion}

The results from both the questionnaires and project outcomes reveal that Lean R\&D was widely accepted by stakeholders at both Petrobras and Americanas, establishing its potential to meet a key success factor in IAC projects: achieving short-term, impactful results for industry. Stakeholders found Lean R\&D's structure conducive to rapid iteration and tangible business outcomes, with many affirming the approach as suitable for aligning academic R\&D efforts with pressing industry needs. 

This compatibility highlights Lean R\&D's utility in fulfilling the short-cycle demands of innovation projects, a benefit that distinguishes it from more traditional R\&D frameworks, which often prioritize long-term objectives over immediate industry impact.

Another key factor contributing to Lean R\&D's success in these projects was the iterative feedback loop between academic researchers and industry practitioners. This continuous feedback cycle allowed both parties to assess intermediate outcomes and make necessary adjustments, enhancing the relevance of the research outputs to address real-world issues effectively. This collaborative approach fostered an environment where theoretical knowledge and practical expertise could merge, enabling more accurate targeting of industry needs and ensuring that the academic efforts aligned closely with the company’s innovation priorities.

However, these positive results should be viewed in light of other contributing factors that are critical to successful IACs \cite{wohlin2011success}. The presence of collaboration champions on-site was particularly advantageous, as they provided essential project oversight and facilitated communication between academic and industry teams. 

Additionally, strong support from company management and alignment with business strategy were crucial. Such backing allowed for smoother integration of academic R\&D outputs into business operations, helping Lean R\&D projects gain traction and relevancy within the company. A shared commitment to achieving industry results, coupled with an innovative research environment at the university, also played significant roles in supporting Lean R\&D's acceptance and perceived effectiveness.

Further, the observed outcomes emphasize the adaptability of Lean R\&D in different settings and its responsiveness to specific project requirements. In the case of Petrobras, the experienced R\&D team leveraged Lean R\&D to achieve measurable business outcomes in AI-driven projects, which led to international patent applications and prestigious awards. 

Moreover, the Lean R\&D approach's adaptability to evolving project scopes and changing business environments proved beneficial. With this respect, in both the Petrobras and Americanas projects, dynamic shifts in priorities due to market demands or organizational needs were seamlessly incorporated, demonstrating the approach's flexibility. This capacity to pivot and reprioritize based on real-time industry feedback ensured that the projects maintained their relevance and impact, further reinforcing Lean R\&D's position as a practical methodology for driving meaningful collaboration in IAC projects.

This not only underscores Lean R\&D's effectiveness in advancing sophisticated technological solutions but also demonstrates its value as a methodology capable of accelerating technology transfer from academia to industry. Petrobras's application of the MVP results in larger-scale development frameworks, like SAFe, further illustrates Lean R\&D's complementary role in innovation pipelines.

For Americanas, involving less experienced students in Lean R\&D projects provided a distinctive perspective on the approach’s versatility. While some students highlighted challenges related to learning and applying the methodology, the majority affirmed Lean R\&D's suitability and appreciated the opportunity to participate in every project phase, from ideation to MVP delivery. This empowered students to engage actively with real-world industry challenges, preparing them for future roles in R\&D and fostering talent trained in the company's specific innovation context.

In summary, Lean R\&D’s ability to deliver impactful results in both short cycles and complex industry settings, combined with its strong acceptance among diverse stakeholder groups, positions it as a valuable framework for IAC projects. Nevertheless, the feedback received reveals opportunities for refinement, particularly in supporting the description of business hypotheses and clarifying the expectations around each phase's activities. By addressing these improvement areas, Lean R\&D can further solidify its role in industry-academia collaborations to produce rapid and business-relevant results.

It should be noted that we report on experiences that, while valuable, inherently lack the scientific rigor associated with more systematic investigations. Specifically, the absence of a formal baseline or direct comparison to alternative frameworks, such as the Technology Transfer Model for co-production~\cite{gorschek2006model,gorschek2021solving}, limits the conclusions drawn about the effectiveness of Lean R\&D in industry-academia collaboration. This shortcoming makes it challenging to contextualize the positive results and evaluate them relative to existing approaches.

\section{Conclusion}

Lean R\&D has been used in IAC innovation projects of the ExACTa initiative with several industry partners. In this paper, we reported on recent results of applying the approach with Petrobras and Americanas, two large companies from the oil and gas and retail industry sectors.

In both experiences, the approach helped bring ideas to impactful business results. The projects delivered to Petrobras resulted in international patent applications and received a Petrobras inventor award at the end of 2022. Their ROI (calculated by the company and considering the investment in the R\&D project against the yearly financial gains obtained from the solutions) exceeds the IAC investment by a factor of more than 1:20. With respect to Americanas, MVPs were successfully delivered in 2023, also positively impacting the business, as the company publicly acknowledged.

We also report perceptions on Lean R\&D from the involved stakeholders. The assessment of the suitability of the phases pointed out that Lean R\&D is perceived to help in defining a joined MVP vision, addressing research-related uncertainties early (in the technical feasibility phase), and efficiently delivering valuable MVPs. Furthermore, we observed the stakeholders' acceptance of the approach.

Opportunities for improvement include providing support for describing business hypotheses and clearly stating the expectations for activities conducted within each phase, making learning and applying the approach easier. Another opportunity one of the sponsors mentioned was more clearly defining the nature of the university collaboration with a greater focus on research, which could bring more innovative results.

Finally, we believe that sharing these experiences presents a valuable contribution to the community. In particular, the adaptability of Lean R\&D across varying project scopes and company needs was evident, showcasing its potential to serve as a flexible methodology for different industry sectors. In the reported experiences both companies benefitted from Lean R\&D's capacity to pivot in response to evolving project requirements and business priorities, ensuring relevance and maintaining momentum toward impactful results. This adaptability further emphasizes Lean R\&D’s strength as a framework for industry-academia collaboration, capable of aligning academic research efforts with dynamic industry landscapes to create practical, high-value outcomes.

%
%
%
\bibliographystyle{splncs04}
\bibliography{mybibliography}

\end{document}